# MULTI-OBJECTIVE RECONSTRUCTION OF SOFTWARE ARCHITECTURE

Frederick Schmidt, Stephen MacDonell
*Software Engineering Research Laboratory, Auckland University of Technology, Private Bag 92006, 55 Wellesley Street East, Auckland 1010, New Zealand*
stephen.macdonell@aut.ac.nz

Andy M. Connor✉
*Colab, Auckland University of Technology, Private Bag 92006, 55 Wellesley Street East, Auckland 1010, New Zealand*
andrew.connor@aut.ac.nz

**Abstract**

*Design erosion is a persistent problem within the software engineering discipline. Software designs tend to deteriorate over time and there is a need for tools and techniques that support software architects when dealing with legacy systems. This paper presents an evaluation of a Search Based Software Engineering (SBSE) approach intended to recover high-level architecture designs of software systems by structuring low-level artefacts into high-level architecture artefact configurations. In particular, this paper describes the performance evaluation of a number of metaheuristic search algorithms applied to architecture reconstruction problems with high dimensionality in terms of objectives. These problems have been selected as representative of the typical challenges faced by software architects dealing with legacy systems and the results inform the ongoing developed of a software tool that supports the analysis of trade-offs between different reconstructed architectures.*

## 1. INTRODUCTION

Contemporary software systems that comprise any reasonable amount of functionality are invariably accompanied by a non-trivial degree of complexity [1]. Additionally, organisations face a steady increase of legacy code that is hard to maintain, highly integrated with other components and therefore very hard to independently refactor or even replace. Furthermore, any given system structure is not static; the structure of the system changes through maintenance, requirements changes, added features and refactorings [2]. This creates difficulties for individuals attempting to understand the design, structures, and dependencies that form the architecture of a software system.

Adding new functionality to an existing software system without considering the conceptual architecture or maintaining the integrity of the software system can result in system erosion. As a consequence software quality decreases and the system will be less flexible, less robust and harder to maintain and understand. Therefore, the software maintenance cost increases. To confine or even reverse system erosion, methods of intensive reverse engineering, manual analysis and refactoring are generally required to re-establish a structured, violation-free and current architectural design. However, development stakeholders often hesitate to engage in such complex and labour intensive tasks due to other pressing commitments and deadlines [3].

This paper outlines a semi-automated approach to re-establish an appropriate architectural design for legacy software systems. This approach is based on the use of multi-objective evolutionary computation approaches in conjunction with software metrics to empower a software architect to select a preferred architecture. Hence, this paper presents also a performance evaluation of a set of six multi-objective evolutionary algorithms (MOEA) in the targeted problem domain. This set of algorithms is not exhaustive, however is reasonable representative of other methods that could or have been applied to this class of problem.

The remainder of this paper is structured as follows. Section 2 outlines some of the related work in this area, whilst Section 3 describes the architecture reconstruction approach. Section 4 outlines the result obtained applying different metaheuristic algorithms to a number of different system evaluations. Finally, Section 5 concludes the paper.

## 2. BACKGROUND AND RELATED WORK

Erosion of software systems is not a new phenomenon and so it has been widely discussed to date. Lehman [4] formulates a set of laws that explain the inevitable and continuous evolution of software systems. These laws express the phenomena of continuing change, steady development, continuing growth, increasing complexity, declining quality and self-regulation. The existence of these laws has been empirically confirmed in a wide range of software system projects [5]. Consequently, it appears that the erosion of software systems is an inevitable side- effect that is likely to become evident in non-trivial software systems.

The increase in complexity combined with an often evident lack of documentation hinders development stakeholders to maintain design aspects of a system [6, 7]. Consequently, uninformed design decisions might impact the architectural integrity of the system. Unhindered deterioration can lead to unsustainable designs, which leave only a complete redesign as a feasible option [3]. However, even if a software system does not become completely inoperative, erosion will make the system more predisposed to defects, high maintenance costs and degrading performance. This in turn is likely to lead to even more erosion. This cycle of erosion can degrade the value, usefulness and technical dominance of a software product.

**2.1. *Design of software architecture***

The architecture of a software system is an abstract model of that system, where fine-grained entities are classified into increasingly abstract modules. An architectural view of a system therefore raises the level of abstraction, hiding details of implementation, algorithms and data representations [8]. Having a current representation of the system architecture is crucial in order to maintain, understand and evaluate a large software application [9]. De Silva and Balasubramaniam [3] state that a good understanding of the physical design of the system and of the flaws in the system is by itself most likely insufficient to prevent the erosion of design if no understanding of the targeted design exist. In this sense, the targeted design can be considered as both the implementation design and a conceptual design [10].

Fowler [11] describes the conceptual design as a view onto the system from a coarse grained and abstract level. The implementation design artefacts are partitioned into coarse grained artefacts such as packages, directories, libraries, subsystems, layers and maybe even layer groups. These coarse grained conceptual artefacts serve as containers in which to accumulate more detailed design artefacts or even finer grained conceptual artefacts from lower conceptual design levels that feature a mutual architectural design attribute [8]. From a conceptual architecture design perspective it is aspired that conceptual artefacts accumulate artefacts that comply with a certain technical, domain or environment aspect [12]. For example, high-level artefacts might accumulate view, client or database functionality. Correspondingly, high-level artefacts should disclose details about the technical implementation or frameworks applied within the application [13].

The conceptual architecture of an application can be modelled to facilitate different architecture patterns [12]. These architecture styles support different domain, quality and environment requirements. Depending on the requirements and purpose of the system, multiple styles are combined to define a complete conceptual architecture model. Murphy, Notkin and Sullivan [14] emphasise that the compliance of the physical architecture and the conceptual model needs to be continuously checked and that the two should be aligned as needed, to obtain a violation-free architecture. Ideally, the physical dependencies should align with the conceptual architecture model of the system [12], although in practice this may not always be achieved. An architecture violation is therefore understood as a dependency within the physical dependency structure which conflicts with the defined dependencies of the conceptual architecture [11]. Identified architecture violations need to be eliminated to reflect the desired architecture design and obtain a code base [11].

The research described in this paper focuses exclusively on the implementation perspective of software architectures. The representation and analysis of software architecture on the implementation and dependency level supports the understanding of the current design of a system and is crucial in enabling the identification of design flaws [15].

**2.2. *Quality assessment of software architecture***

The presence of erosion can be inferred by considering metrics determined from the software system and so there is potential to use appropriate software metrics to direct a semi-automated approach to software architecture reconstruction to reverse or limit software erosion. The research described in this paper implements a search based driven software modularisation approach. The concept suggested by Harman and Clark [16] to utilise software metrics as a fitness function is implemented in this research to evaluate the generated architecture configurations and enable the navigation through the search space. To that end, different software metrics can be considered as measures of quality of a software architecture or a software system.

There is general consensus within the software engineering community that high cohesion within artefacts and low coupling between artefacts is a desired design goal and an indicator of good design [13]. The benefit of such a design is that artefacts that feature high cohesion are easier to understand, test and maintain. Various approaches to measure cohesion and coupling on compilation unit level have been applied in single objective optimisation approaches to guide the search process [17, 18]. Gui and Scott [19] suggested to measure cohesion on higher abstraction levels to evaluate component re-usability, and as such they defined the cohesion of a high-level artefact as the mean of the cohesion measures of all members of the artefact. Given the focus of this work, cohesion metrics that operate on higher abstraction levels such as package, subsystem and system level are relevant in terms of addressing the objective of the present research.

Coupling measures the strength of dependency between artefacts [20]. Consequently, coupling gives an indication to what degree a software artefact relies on each one of the others. Low or loose coupling indicates that the source code is organized in such a way that it features no strong dependencies between each of its members [21]. Design techniques to achieve low coupling within software systems are, for example, prevention of cyclic dependencies, maintenance of high-level dependency structures, referencing of interfaces instead of concrete types and application of dependency injection frameworks. The existence of cyclic dependencies limits reusability, testability and the impact analysis of changes in the involved system artefacts. Empirical evidence supports that cyclic dependencies are evident in almost all non-trivial software systems on lower abstraction levels [22]. Recent research empirically underpins that most cycles on compilation-unit level do not deteriorate the testability and reusability of software systems [23]. Falleri, Denier, Laval,

Vismara, and Ducasse [24] argue that the composition of individual cycles should be considered, to assess the impact on the quality of a software system. For example, longer cycles have a more negative impact on the structural quality of a software system. Hence, an established principle of good architectural design is that software architectures feature a cycle-free design on package and higher abstraction levels to support quality attributes such as testability, reusability and understandability [13]. Furthermore, Lakos [10] expresses the quality of a solution based on dependency characteristics of the system through the use of the *Normalised Cumulative Component Dependency (NCCD)* metric. The metric suggests that the components of system should be organized using a balanced binary model. This organization can lead to high reusability, good analyzability and testability. Thus, the optimisation of architecture compositions towards designs that feature a low number of cycles and binary tree like structures is a worthwhile objective and hence pursued in this research.

The research described in this paper addresses the reconstruction of a software architecture design by classifying low level-artefacts (e.g. compilation units and/or packages) into artefacts of higher abstraction levels (e.g. packages, subsystems and layers). Previous research in this area focused on the grouping of elements into the next highest abstraction level [18, 25, 26]. When dealing with different abstraction levels it is worth considering the structure of the elements (e.g. desired dependency flow, cycle free organisation) within high-level artefacts and to discuss the impact on quality aspects of the reconstructed architecture.

### 2.3. Search based modularisation

This paper presents an approach that applies MOEA implementations in architecture reconstruction. This approach is implemented as a software modularisation approach that considers a conceptual architecture model during the modularisation process. Whilst a number of approaches have been described in the literature that involve the discovery of an architecture [27, 28], the work in this paper differs in that a conceptual architecture is applied and existing compilation units mapped into the architecture. This builds on various approaches that implement search based techniques within low-level software modularisation. Mancoridis, Mitchell, Chen and Gansner [29] and Seng, Bauer, Biehl and Pache [30] present the main approaches to applying SBSE techniques to reengineer the structure of a software system.

Mancoridis et al. [29] show that the structure and complexity of cluster analysis as applied to software systems is a promising approach to create feasible solutions. The objective of the approach is to discover a cluster configuration which features high cohesion within clusters and low coupling between clusters and has been extended by Mitchell [17], Mitchell and Mancoridis [18] and Mitchell and Mancoridis [31]. Other work in this areas has included the evaluation of different metrics for using in the clustering approaches and the implication for the robustness of software modularisation [32].

Abdeen, Ducasse, Sahraoui, & Alloui [25] apply Simulated Annealing to optimise class partitioning within the existing package structure of a software system. Seng et al. [30] also present a single objective approach that searches for an optimal subsystem decomposition by optimizing metrics and heuristics of good subsystem design. This approach groups compilation units into a higher abstraction level. From a software design perspective, the subsystems can be understood as packages or folders of the software system. Similarly, Schmidt, MacDonell and Connor [33] present a feasibility study of the application of an automatic refactoring approach to increase cohesion of packages, reduce coupling between packages and reduce the number of architecture violations in the model of a software system by source code refactoring on that enabled the resolution of dependencies on compilation unit level.

Etemaadi and Chaudron [34] propose a conceptual framework for the application of multi-objective optimisation for the design of embedded architectures. Additionally, they highlight NSGA-II and SPEA2 as promising algorithm candidates for the implementation of such a framework. Praditwong et al. [35] also approach software clustering from a multi-objective perspective by implementing the concepts of Pareto Optimality and Non-Dominance. This work highlights that most previous studies only utilised fixed weighted agglomerations of high cohesion and low coupling as a single objective function as opposed to a true multiobjective approach. Barros [36] extends this line of thinking and compares the performance of a multi-objective clustering approach with three different objective configurations.

The research described in this paper was completed in 2014, since that time other research has emerged related to multiobjective remodularisation of software systems. For example, Mkaouer et al. [37] utilise the NSGA-III algorithm on the remodularisation of a number of different software systems. Whilst this work utilises software metrics as the foundation of the fitness function, the research is based around determining the efficiency of the approach which differs from the work outlined in this paper. The Rearchitecturer system is a practical tool that allows filtering of solutions across any number of multiple objectives to allow a software architect to inspect multiple potential solutions. The Rearchitecturer tool itself is not described in this paper, but is available as an open source solution[1] for researchers to access.

In general, interesting techniques and approaches have been proposed to overcome challenges in the application domain of search based software modularisation. However, there is still considerable scope to consider how MOEA implementations can be applied to assist software architects with the challenges of dealing with software erosion.

---

[1] https://sourceforge.net/projects/rearchitecturer/

# 3. ARCHITECTURE RECONSTRUCTION APPROACH

A prototype, namely the Rearchitecturer system, has been designed and developed to enable the reconstruction of software architecture configurations based on the application of multi-objective optimisation techniques. The objective of the system analysis and evaluation stage is to gather data that enables the formulation of conclusions on the feasibility, contributions and limitations of the developed system.

### 3.1. Reconstruction scenarios

The present research applies the concepts presented in Harman and Clark [16] that outline the application of software metrics as fitness functions to determine the fitness of a generated solution. The approach developed in this research facilitates the classification of software artefacts into conceptual architecture models on multiple abstraction levels. Additionally, conceptual architecture models can be reconstructed as part of the reconstruction process. However, following Breivold, Crnkovic and Larsson [38], Fowler [11] and Martin [1], this research considers the conceptual architecture model as a blueprint of the desired design of the system. A conceptual architecture model is supposed to describe the design of the system based on domain aspects [1]. Hence, the design of the conceptual model is ideally driven by domain aspects and their relationships. The implementation of the system should adapt to this blueprint. Correspondingly, the conceptual model should not reflect the implementation of the system. Hence, from an architecture design perspective the rebuilding of the conceptual architecture based on the structures of the physical source code artefact conflicts with the ideas presented in the mentioned architecture design literature. Therefore, experiments are conducted in the evaluation of this research that consider predefined conceptual architecture models. These conceptual architecture models operate as target architectures during the classification process.

The conceptual architecture model can be utilised to model different system architectures. For example, a transient architecture style can be utilised if the application is running on one machine and no machine boundaries exist [11]. A strict layer dependency, in which a layer can only access artefacts in the directly depending layer, can be applied to model machine boundaries [11]. The structure of the conceptual architecture is considered by some of the employed architecture design metrics. Hence, the employment of different architecture styles within the evaluation is useful to reveal information on the applicability of the approach within different architecture styles.

Principles of good architecture design envisage that classification of physical source code artefacts into conceptual high-level artefacts is driven by the intended functionality of the physical artefacts [1]. The optimisation implemented in this research is based on established architectural design indicators and does not automatically consider functionality aspects of the physical artefacts. However, stakeholders might understand this intended functionality for some of the physical artefacts that they want to have included in the solution. Hence, stakeholders can assign artefacts prior to the execution of the search to include their domain knowledge into the architecture reconstruction process by assigning class or package patterns to certain high-level artefacts. The developed approach considers such predefined assignments. As a result, any visited solution will feature the predefined artefact assignments. The consideration of such a manual assignment impacts the solution space of the search and impacts the performance of the employed search. Hence, the inclusion of an experiment scenario that considers predefined artefact assignments is helpful to gather data on the performance of the employed MOEA implementations.

However, the presented results focus on an experimental set in which the feasibility of a randomised assignment of source code artefacts combined with the subsequent execution of MOEA implementations in constraint classification scenarios is evaluated. It is necessary to conduct multiple iterations of the randomised assignment of physical source code artefacts and the subsequent optimisation to gain a set of results that enables representative conclusions.

### 3.2. Evaluation systems

The defined problem scenarios and MOEA configurations are applied on a set of software systems to enable conclusions to be drawn on the performance, applicability and scalability of the developed approach. The approach has been evaluated on the following five software systems: Apache Log4j, Apache Commons Math, Apache Ant, Lucene and the Rearchitecturer system itself. These systems (other than Rearchitecturer) were chosen because they exhibit the following characteristics.

All the systems are established open-source projects with an active user community. Multiple developers are permanently involved in the maintenance and enhancement of these projects. These systems feature a module-based architecture that is publicly available on the corresponding system webpages. The modules in these systems are maintained as separate projects. The dependencies between the modules are organised with Maven. Hence, at a project level the systems are cycle free and have a defined dependency structure. The systems also represent different sizes and complexity that potentially provides insight as to how a MOEA reconstruction approach might scale to different applications. However, it is noted that all the systems are generally on the small side when compared to the full set of possible software systems that could be used.

In contrast, the Rearchitecturer system has been developed as a prototype for the evaluation of this research. Thus, one developer was involved in the development of the Rearchitecturer system and no active user-community exists at this stage. It is a completely known system in which the intended architecture is both defined and understood.

The selected systems are of different size, structure and maturity. Table 1 depicts software metrics that describe aspects of the size and structure of the selected software systems.

Table 1. Size metrics of evaluation systems

| Name | No. Packages | Lines of Code | No. Types |
|---|---|---|---|
| *Apache Ant v.1.9.2* | 276 | 131,212 | 1,772 |
| *Apache Math v.3.2* | 140 | 171,171 | 2,106 |
| *Apache Log4j v.1.2.17* | 40 | 30,456 | 453 |
| *Lucene v.4.4.0* | 50 | 151,340 | 2,264 |
| *Rearchitecturer* | 51 | 33,231 | 537 |

### *3.3. Selected Algorithms*

This research investigates the performance of a set of six MOEA (AbYSS, GDE3, MOEAD, NSGAII, OMOPSO, Random) implementations across the five different software systems described in the previous section. The implementations of the JMetal MOEA framework are utilised in the present research and details of the specific algorithms are available in the description of the JMetal framework [39]. The selection of algorithms is intended to be representative of parallel search methods but not intended to be an exhaustive study of all available algorithms.

### *3.4. Optimization Objectives*

The Rearchitecturer system features a large number software quality metrics that can be considered based on the needs of the development stakeholders. However, only a small number of these metrics are used in this paper as an example of the capabilities of the approach. The selection of optimization objectives has been driven with a strong focus on best software engineering practices as discussed in the literature review. The following set of eight selected optimisation goals is utilised in this research:

1. Relational cohesion in subsystems (maximise)
2. Normalised cumulative component dependency of subsystems (converge to 1.0)
3. Efferent coupling of subsystems (minimise)
4. Afferent coupling of subsystems (minimise)
5. Distance in subsystems (minimise)
6. Number of forbidden outgoing type dependencies (minimise)
7. Number of package cycles (minimise)
8. Range of compilation units in subsystems (minimise)

An exception to the strategy of selecting best practice metrics is the range of compilation units in subsystems metric. It has been found through the course of this research that there was a tendency of the optimisation approaches to organise low-level artefacts into a small number of high-level artefacts to reach good performance in the number of cycles on package level, number of architecture violations on compilation unit level and coupling metrics. The other metrics have not been able to counteract this movement. As a result, the employment of the original seven software engineering optimisation goals led to an unacceptable number of solutions that featured an organisation of most low-level artefacts in only a few big high-level artefacts. The range of compilation units in subsystems metric has been introduced to counteract this tendency. However, it is acknowledged that the minimisation of the range of compilation units in subsystems metric is not an architecture design metric that would usually be seriously considered in the manual design process of software architecture configurations.

Whilst it is noted that the relatively large number of objectives can produce poor performance, this research purposefully sets out to consider problems that are representative of the typical challenges faced by software architects attempting to improve the quality of legacy software systems. A more detailed performance analysis of the algorithms on a smaller set of objectives could be undertaken as future work.

To thoroughly evaluate the performance of a multi-objective approach, all the employed objectives need to be taken into consideration. Hence, the performance assessment of multi-objective based optimisation approaches is more complex. The major difficulty of multi-objective assessment is that the output of the optimisation process is not a single solution but rather a non-dominated Pareto-Front. The performance evaluation of the algorithms should consider both the ability to converge to a set of solutions that exhibit desirable objectives but also how representative the Pareto-Front is of the solution space. This can be undertaken using performance indicators to assess the quality of Pareto-Fronts [40-42]. Such performance indicators do not provide any insight to the quality of solutions found in the objective (fitness) space, but do provide insight into the performance of the algorithm in terms of both exploring the solution space and convergence.

Generally, performance indicators that assess convergence (e.g. Generational Distance, Epsilon Indicator) and diversity (e.g. Spacing) aspects of Pareto-Fronts can be differentiated. However, hybrid forms, which express convergence and diversity aspects in one metric exist [43] (e.g. Hypervolume). Some performance indicators require the true Pareto-Front in order to determine a value. The true Pareto-Front is understood as the best achievable Pareto-Front of a problem [44]. Often the true Pareto-Front is not known or cannot be calculated for a problem. This certainly applies for the research presented in this paper, as the true Pareto-Front, that entails all non-dominated architecture configuration solutions for a selected optimisation configuration, is not known and cannot be generated deterministically. The developed evaluation approach creates a super Pareto-Front from the individual optimisation runs that can be considered an approximation of the true Pareto-Front. The creation of such a super Pareto-Front is an established technique of evaluation and has been applied in other optimisation research [36, 45, 46]. A limitation of utilising an approximation instead of the true Pareto-Front is that the performance indicator only calculates a relative convergence and diversity. Hence, results that have been created with different super Pareto-Fronts are not directly comparable. In the present research this needs to be considered if Pareto-Fronts are calculated with different objective settings or with different software systems. Hence, an approach is suggested that relies on the application of normalisation techniques to enable the comparison of performance indicator results from such incompatible Pareto-Front calculations.

## 4. RESULTS

This section demonstrates the application of the developed multi-objective evaluation framework in the problem domain of architecture reconstruction. Three dimensions are considered

### 4.1. *MOEA Performance in Multiple Architecture Reconstruction Scenarios*

In this research, a pre-defined conceptual architecture model is considered and so the conceptual architecture is not discovered during the reconstruction process. Correspondingly, the reconstruction configuration employs the assignment of compilation units into the existing packages of the system and the assignment of the packages into the layers of the conceptual target architecture.

Experiments with transient and strict conceptual architecture models with a different number of layers (2, 3 and 4 layers) have been conducted in this research. The general structure of these conceptual models is based on the C2-architecture-style to support separation of concerns and high-level modularisation of the reconstructed system [12].

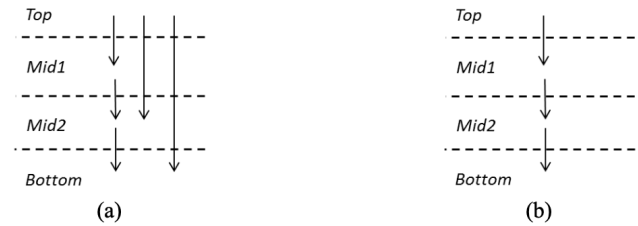

Fig. 1. Transient and strict architecture models

As expected it has been found that the application of conceptual architecture models with a higher number of layers is a more complex search problem. Consequently, the absolute achievement of MOEAs in conceptual architecture target models with a lower number of layers is better. Nevertheless, it has been found that the application of a different number of conceptual architecture layers does not impact the individual MOEA implementations in comparison to one another. Hence, in the experiment presented in this section only conceptual architecture models that feature four conceptual layers are employed. However, three different conceptual architecture paradigms are used to evaluate the performance of MOEAs in different architecture reconstruction scenarios.

The first conceptual target architecture model features four transient layers in which each top layer can access any bottom layer. Figure 1(a) depicts the employed transient architecture model. Such an architecture design is common when no distribution of the system is evident and consequently all the artefacts are available on the same machine.

The second conceptual target architecture model features four strict layers in which each top layer can access only one directly depending layer. Figure 1(b) depicts the employed strict architecture model. Such an architecture design can, for example, be used to model a distribution of layers across machine boundaries. Displayed equations should be numbered consecutively in the paper, with the number set flush right and enclosed in parentheses.

In the third reconstruction scenario, the strict target conceptual architecture as presented in Figure 1b is utilised and at the beginning of each seed the packages of the optimised software system are assigned randomly to each layer. The compilation units that are included in the corresponding packages are not reassigned. Hence, every solution features the initial package assignment of the corresponding seed. The idea of the chosen start configuration is to simulate that the original compilation unit configuration represents a fairly good solution that ideally needs only a fine-grained reorganisation instead of a complete reassignment of packages and compilation units. Additionally, development stakeholders might have a reasonable understanding of the quality of the original assignment of some packages and would like to include this knowledge in the created solutions.

The *MOEA* implementations that are considered enable the tuning of a variety of optimisation parameters (e.g. population size, number of iterations, different crossover and variation operators and the corresponding crossover and mutation parameter settings). It has been found within the tuning phase that neither population size, mutation and crossover parameters changes the search outcome significantly. A population size of 50 and the following variation operator settings are applied in the remaining experiments to enable the traceability of the presented experiment results: mutation rate= 0.5, mutation distribution index = 10.0, crossover rate = 1.0 and crossover distribution index = 10.0. Experiments have shown that good convergence of the applied *MOEA* implementations is usually achieved within 3,000 – 5,000 iterations. That said, 50,000 iterations are used as a termination criterion in the following experiments. A total of 90 different experiment configurations (6 MOEA implementations x 5 systems x 3 target architecture reconstruction scenarios) are executed to gather the data presented in this research. Each search configuration is executed 10 times to accommodate the probability characteristics of the MOEA implementations. Hence, the described experiment configuration features a total of 900 solution sets.

Search definitions which describe the applied MOEA configuration, software system, target architecture and number of seeds are generated by the Rearchitecturer system. A search executer schedules the execution of individual search seeds on designated worker nodes. Hence, a worker node executes a seed at a time. The optimization result set is streamed back from the worker node to the master after the execution of the seed. A statistical analytics engine has been implemented through the course of this research that is able to query the consolidated search results. The evaluation engine allows the calculation of describe statistics of objective and performance metric achievement based on defined configuration parameters. For example, algorithm performance can be analysed across different parameter tunings, software systems and target architecture designs. The classification of the evaluation systems into the different conceptual target architectures features different levels of complexity. For example, the resolution of architecture violations is harder within a strict architecture than the classification into a transient target architecture. Additionally, the solution space is constrained in the reconstruction scenario, in which packages are fixed into layers. Hence, finding solutions that feature good performance in the subsystem structure metrics is more complex. To demonstrate the impact on the different target architectures the dataset is sliced by consolidating algorithms and systems and the separation of result sets is based on the different target architecture designs. The analysis of the sliced datasets confirmed a different level of performance achievement depending on the level of the complexity of the reconstruction scenario. This can be demonstrated by considering how the Non-Dominated-Pareto-Front (NDPF) develops by considering all of the collected data split by reconstruction scenario.

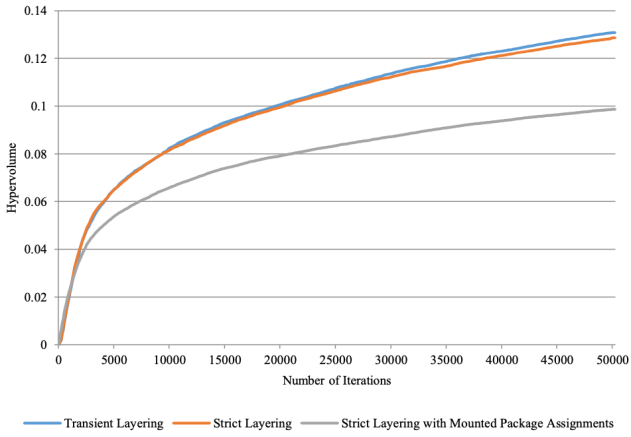

Fig. 2. Development of hypervolume

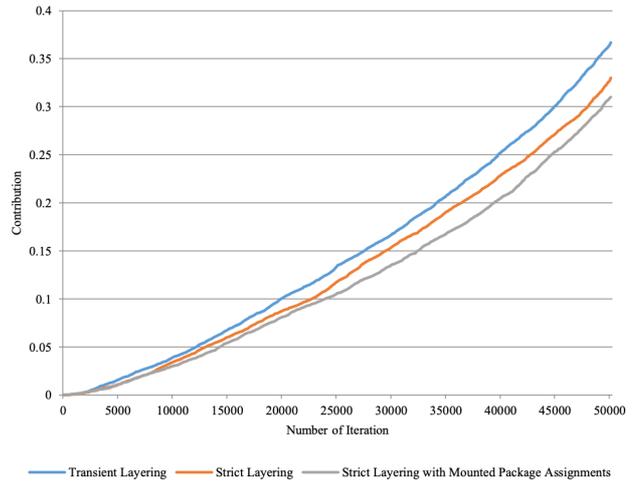

Fig. 3. Development of contribution

Figure 2 depicts the corresponding Hypervolume performance development graph for the three cases of transient layering, strict layering and strict layering with mounted packet assignments.

The graph shows that the transient- and strict-reconstruction scenarios achieve equally good Hypervolume performance. Understandably, the Hypervolume performance of the third architecture example is substantially lower due to the increase of complexity based on the upstream assignment and mounting of packages into subsystems.

In the results relating to objective space presented later in this paper, the performance of the employed MOEA implementations across the three reconstruction scenarios is evaluated. The three architecture reconstruction datasets are consolidated to enable such an analysis. However, a prerequisite for the validity of such a comparison is that each reconstruction scenario contributes to the approximated true Pareto-Front, as the analysis relies mainly on achievement in the objective space and convergence of the optimal Pareto-Fronts. The solution sets of a reconstruction scenario are excluded from the analysis if the optimal Pareto-Front of that reconstruction scenario is not contributing to the approximated true Pareto-Front and the slicing is based on the employed MOEA implementations. For example, if the least complex architecture reconstruction scenario dominates the optimal Pareto-Front of the more complex architecture scenarios the analysis only considers the solution sets of the least complex architecture reconstruction solution sets. The Contribution performance metric is useful in this regard to confirm if each reconstruction scenario can contribute towards the approximated true Pareto-Front. Figure 3 depicts the development of the contribution performance metric through the search process in the individual reconstruction scenarios.

The contribution graph depicts that the complexity of the search impacts the Contribution outcome. However, the optimal Pareto-Fronts of the individual reconstruction scenarios feature a similar contribution to the approximated true Pareto-Front despite the different complexity evident in the individual reconstruction scenarios. Hence, it can be concluded that the analyses of the MOEA implementations is representative for all three employed reconstruction scenarios.

### 4.2. *MOEA performance in the objective space*

The main objective of this research is the application of multi-objective optimisation techniques in the application domain of architecture reconstruction. High-level architecture design metrics are employed as objectives to implement this research project. Such high-level architecture design metrics have not been employed in related research efforts. The review of the capability of the employed optimisation techniques to advance the individual objective dimensions is necessary to enable statements on the applicability of the selected high-level architecture design metrics to be made.

No generally accepted method has been established in other multi-objective research to assess the achievement in the objective space. Within the present approach the explicit reliance on the application of multi-objective performance metrics that calculate relative convergence of similarity to a best optimal Pareto-Front might be misleading. For example, two optimisation configurations might reduce the number of architecture violations to 500 and 800. A convergence-based performance metric will confirm a better performance for the approach that achieved 500 architecture violations if we ignore the existence of other solutions and objectives in this example. However, both solutions are most likely still too complex to allow stakeholders to understand the solution and manually resolve remaining forbidden dependencies. Hence, both configurations would be infeasible for use in the target application domain. Hence, assessment of the objective achievement is an important component to assess the overall feasibility of the developed approach.

Theoretically, the objective achievement can be assessed based on any of the solution sets evident in a multi-objective optimisation. These kinds of solutions sets are: the complete set of visited solutions, the current population, and the optimal Pareto-Front. The implemented evaluation framework supports the assessment of the objective achievement based on all three of these solution sets. Descriptive statistics are calculated for the individual solution sets at each performance snapshot. The selection of the solution set and the descriptive statistics depends on the user's research interest.

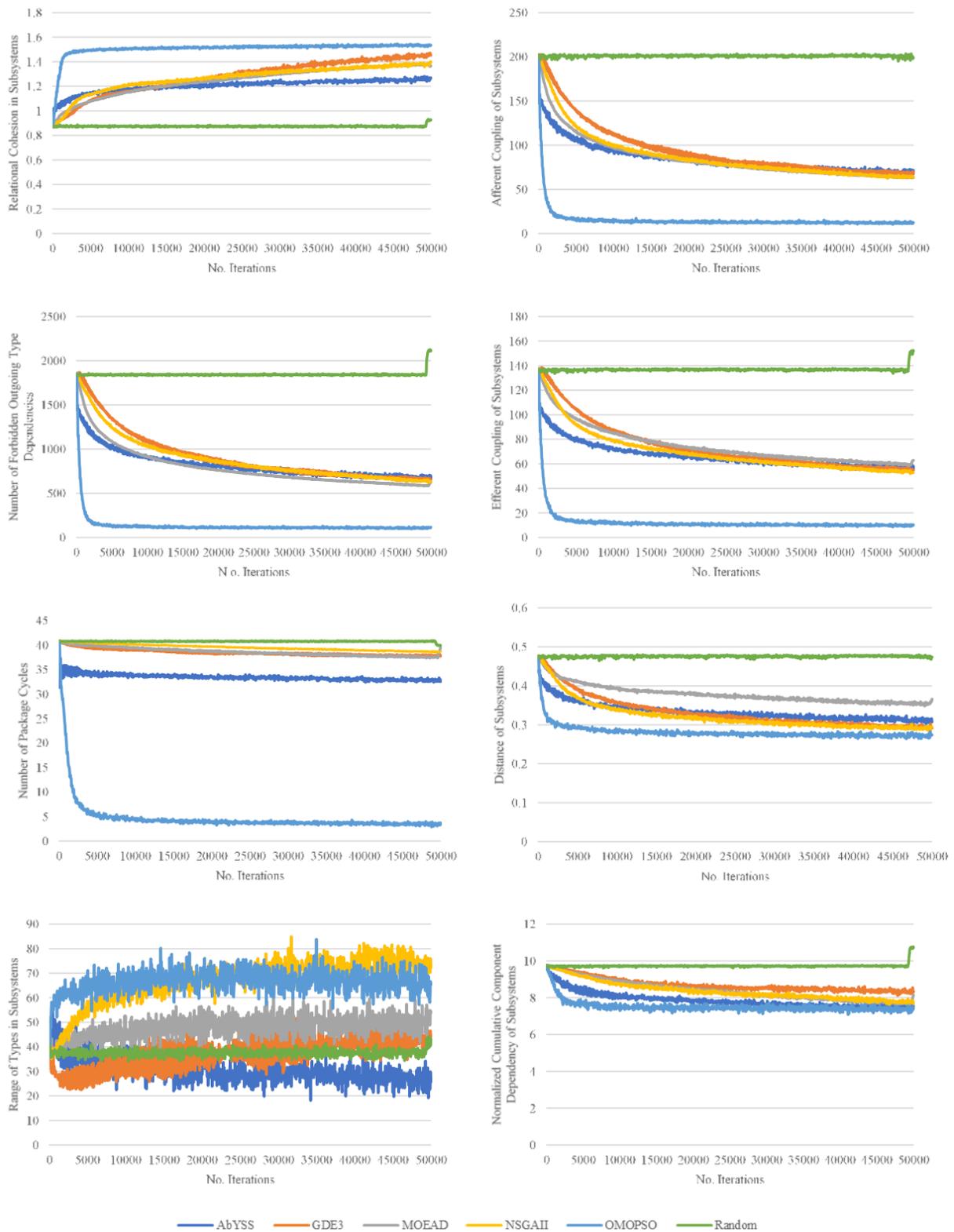

Fig. 4. Objective achievement in populations for the eight objectives

The first analysis demonstrates the convergence in the individual objectives. Hence, the solutions of the generational populations or more specifically the solution that features the best performance in the desired objective is used.

This evaluation uses the same data configuration as described earlier with only one variation operator setting. The results are sliced by the applied MOEA implementation. Correspondingly, a total of 150 (900/6) seeds are considered to determine the performance for each MOEA implementation at each performance snapshot. Multiple runs of the algorithm ensure that any variability in performance that arises from the stochastic nature of the algorithms is mitigated. The evaluation framework reports descriptive statistics for each performance snapshot. In this evaluation, the interval for performance snapshots is aligned with the population size of the generations and is correspondingly set to 50 iterations. Hence, based on the 1,000 generations that are conducted in the experiment setup also a corresponding set of performance snapshots is calculated for this analysis.

Figure 4 depicts the mean development of the best achievement of the eight employed objectives. The mean for each objective is presented for each performance snapshot.

The objective development depicts that each of the employed MOEA implementations features better performance than Random in seven of the eight objective dimensions. Additionally, OMOPSO features the best overall advancement in the same seven objectives. Furthermore, no specific performance differences can be reported between NSGA-II, GDE3, ABYSS and MOEAD in these seven objective dimensions. It has been noted in the literature that MOEA performances rapidly degrades with increasing dimensionality in terms of the number of objectives [40], so improved performance in most algorithms could be achieved through reducing the number of objectives. However, the intention of this research is not to provide an optimal set of solutions but instead to provide potentially feasible options for a software architect to explore and choose between. An exception is the "range of types in subsystems" objective in which only AbYSS shows better convergence than Random. However, it needs to be considered that the presented results are the mean of the best performances per population at each generation of the search. Hence, it is not a representation of the overall search process or achievement in that objective. The presented results are simply an indication of the feasibility of the approach to converge the individual objectives. It may be suggested that the review of the objective achievement of the optimal Pareto-Front is a better evaluation instrument to assess the general feasibility of an optimisation approach in the individual objective dimensions. However, the presentation of the mean development of the objective progress of the optimal Pareto- Front solution set is most likely also an unreliable means to assess the performance of a configuration setting. The reason for this is that the trade-off concept of optimal Pareto-Fronts leads to the inclusion of solutions that dominate any area of the objective space. These solutions therefore might feature poor performance in the reviewed objective. Hence, while a good progress in the minimum and maximum value of an objective is achieved the mean progress might be relatively constant in the reviewed objective. Hence, it is suggested in this research that the reporting of descriptive statistics, and in particular the review of minimum, maximum and distribution characteristics of objective achievement of the optimal Pareto-Front, are more valuable methods to review achievement in the objective space.

This analysis relies on achievement in the objective space in the optimal Pareto-Fronts. The normality condition is not fulfilled in these datasets. Hence, no valid conclusions can be drawn on the distribution of the data based on the Mean and SD measures. Consequently, this paper only considers the Minimum, Maximum and Median of the objective measures of the optimal Pareto-Fronts. A full analysis of all of the objectives is beyond the scope of this paper, hence only one objective is considered as an exemplar of the analysis process.

The populations for these descriptive measures are created based on the objective measures of the solutions of the optimal Pareto-Fronts of the created slices. The results presented in the tables are sliced by the applied MOEA implementation and system to enable the assessment of the objective performance of MOEAs in the individual systems.

Table 2. Descriptive statistics: Number of forbidden type dependencies

| System | Algorithm | Min | Max | Median |
|---|---|---|---|---|
| *Apache Ant* | AbYSS | 460.00 | 3000.00 | 2033.37 |
| | GDE3 | 320.00 | 3000.00 | 1784.63 |
| | MOEAD | 220.00 | 3000.00 | 1524.63 |
| | NSGAII | 500.00 | 3000.00 | 1770.56 |
| | OMOPSO | 1.00 | 3000.00 | 1656.58 |
| | Random | 2000.00 | 3000.00 | 2711.85 |
| *Apache Math* | AbYSS | 11.00 | 19.00 | 16.46 |
| | GDE3 | 13.00 | 19.00 | 17.49 |
| | MOEAD | 12.00 | 19.00 | 17.44 |
| | NSGAII | 13.00 | 19.00 | 17.54 |
| | OMOPSO | 0.00 | 19.00 | 12.57 |
| | Random | 15.00 | 19.00 | 17.64 |
| *Apache Log4j* | AbYSS | 660.00 | 3000.00 | 2119.18 |
| | GDE3 | 330.00 | 3000.00 | 1953.53 |
| | MOEAD | 600.00 | 3000.00 | 1909.11 |
| | NSGAII | 400.00 | 3000.00 | 1910.24 |
| | OMOPSO | 1.00 | 3000.00 | 1458.10 |
| | Random | 0.00 | 3000.00 | 2792.83 |
| *Lucene* | AbYSS | 2.00 | 2500.00 | 2500.00 |
| | GDE3 | 1200.00 | 2500.00 | 2500.00 |
| | MOEAD | 70.00 | 2500.00 | 2500.00 |
| | NSGAII | 850.00 | 2500.00 | 2500.00 |
| | OMOPSO | 1.00 | 2500.00 | 2500.00 |
| | Random | 980.00 | 2500.00 | 2500.00 |
| *Rearchitecturer* | AbYSS | 130.00 | 930.00 | 930.00 |
| | GDE3 | 10.00 | 930.00 | 930.00 |
| | MOEAD | 140.00 | 890.00 | 890.00 |
| | NSGAII | 6.00 | 920.00 | 920.00 |
| | OMOPSO | 1.00 | 900.00 | 900.00 |
| | Random | 580.00 | 910.00 | 910.00 |

Table 2 presents the numerical data relating to the pareto-optimal solutions found by each algorithm for a single objective, the number of forbidden type dependencies.

In general, the presented statistics of the optimal Pareto-Fronts of this objective are similar to those achieved for the other objectives. In this case, OMOPSO features the best absolute achievement in this objective across all systems. An exception is Random that discovered a solution with 0.0 forbidden type dependencies in the apache math system. Additionally, NSGAII and GDE3 can find promising solutions in the Rearchitecturer system and AbYSS is able to find promising solutions for the lucene system.

The other MOEA algorithms show worse performance in more complex systems apart from this coincidental discovery by the Random implementation in the apache math and log4j systems. However, each algorithm has shown the potential to contribute pareto-optimal solutions.

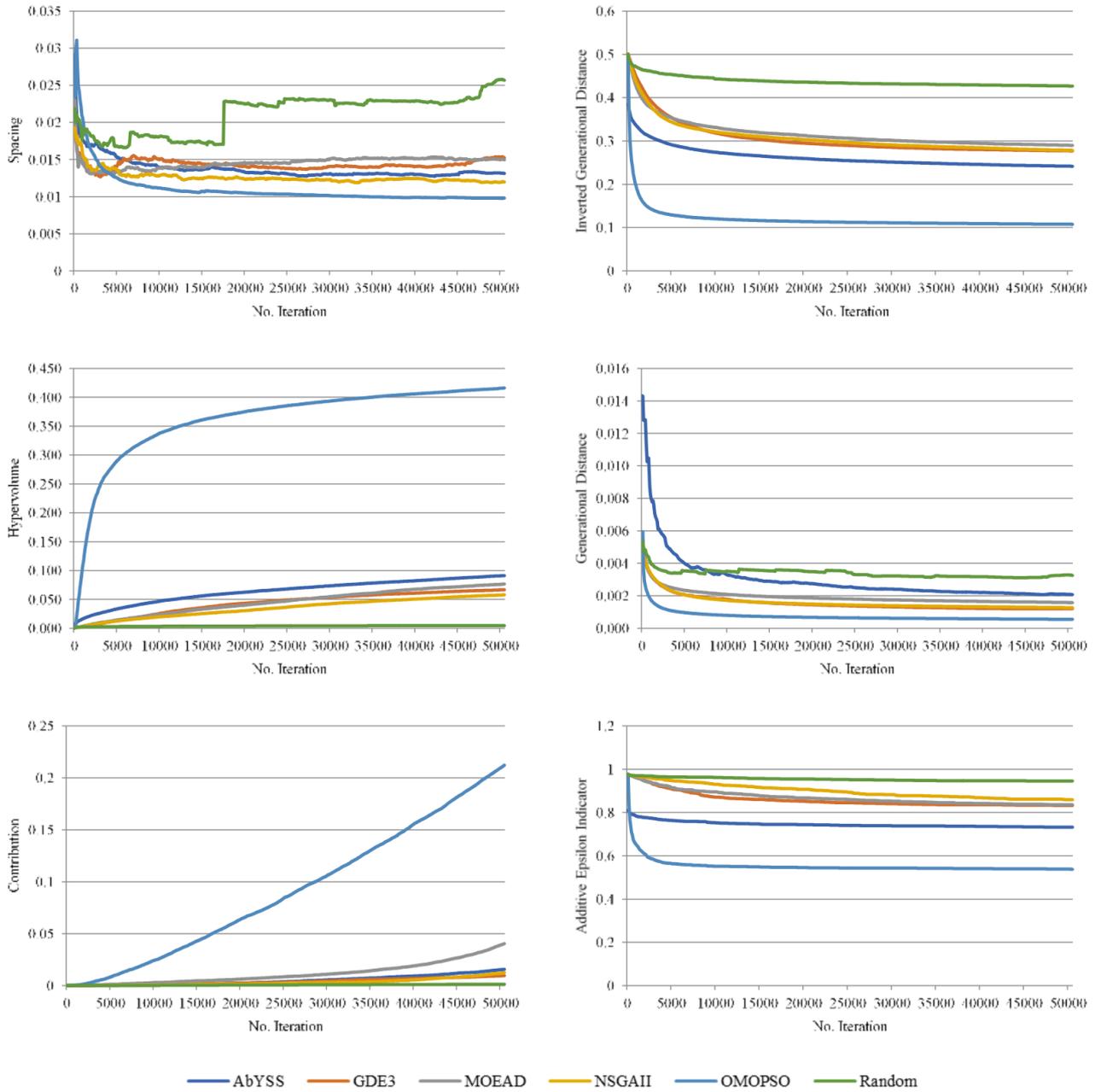

Fig. 5. Performance Indicators - Slicing based on MOEA implementation

Whilst the reconstruction scenario described in this paper utilised a total of 8 objectives, only a single case presented though this is representative of the full set [41] and sufficient to show that all of the MOEA implementations are capable of addressing the reconstruction of the software architecture and find solutions that are not only pareto-optimal but also demonstrate improved performance in the objective space. The approach therefore shows promise in terms of its ability to address the issue of software erosion.

**4.3. *Performance metrics analysis of MOEA algorithms***

The analyses based on descriptive statistics for the individual objectives is first of all complex but also not representative to assess the overall performance of a set of search configurations. For example, differences in performance would occur in the individual objectives and forming a general conclusion on the performance of the individual configuration slices is very difficult based on the statistical analysis of the individual objectives. A more appropriate method to determine the overall performance is the consideration of multi-objective performance metrics. As discussed previously Pareto-Front performance metrics provide a useful means by which to consolidate the performance of multi-objective metrics into a single comparable value. Multi-objective performance metrics such as Hypervolume and Contribution have been presented in earlier sections to reveal performance differences in multiple architecture reconstruction scenarios. Nevertheless, the consideration of multi-objective performance metrics in combination with measures of statistical difference testing are a powerful and straightforward approach to determine and statistically justify the performance differences of multiple search configurations. Hence, the suggested approach is also applied to determine the absolute difference of performance of the employed *MOEA* implementations. Figure 5 depicts the development of the six captured performance metrics based on the discussed dataset and the slicing into the employed *MOEA* implementations.

Table 4. Mean performance of MOEA implementations (Iteration 50,000)

| Performance Indicator | ABYSS | GDE3 | MOAD | NSGAII | OMOPSO | Random |
|---|---|---|---|---|---|---|
| *Spacing* | 0.0131 | 0.0152 | 0.0150 | 0.0120 | 0.0098 | 0.0257 |
| *Inverted Generational Distance* | 0.2621 | 0.3047 | 0.3164 | 0.3075 | 0.1189 | 0.4374 |
| *Hypervolume* | 0.0912 | 0.0666 | 0.0765 | 0.0577 | 0.4162 | 0.0047 |
| *Additive Epsilon Indicator* | 0.7453 | 0.8581 | 0.8690 | 0.8992 | 0.5526 | 0.9538 |
| *Contribution* | 0.0051 | 0.0039 | 0.0115 | 0.0033 | 0.0910 | 0.0007 |
| *Generational Distance* | 0.0021 | 0.0012 | 0.0016 | 0.0013 | 0.0006 | 0.0033 |
| *N* | 180 | 180 | 180 | 180 | 180 | 180 |

All but the *Spacing* performance metric confirm that all *MOEA* implementations feature better performance than a *Random* search. It cannot be determined in this research if a wider *Spacing* is a desired *Pareto-Front* attribute in the targeted problem domain. Additionally, the mean development of performance metrics shows that *OMOPSO* outperforms all other *MOEA* implementations in the presented performance indicators. In the performance metrics that measure relative convergence (*Additive*

*Epsilon Indicator*, *Generational Distance, Hypervolume* and *Inverted Generational Distance*) an almost full convergence can be observed between iteration 2,000-3,000. By comparing the data in Figure 4, Table 2 and Figure 5 then it appears that the OMOPSO algorithm is performing well both in terms of the objective space (as measured by the objective values) and the solution space (as measured by the Pareto-Front performance indicators). No major differences in performance development can be observed that depart from the final performance outcome of the *MOEA* implementation slices. Hence, the remaining analysis focuses only on the discussion of the performance snapshot at iteration 50,000. Table 4 depicts the mean performance indicators of the performance snapshot at iteration 50,000.

The *Kruskal-Wallis* test has been applied due to the departure from normality in all populations in the performance snapshot at iteration 50,000. The pair-wise significance comparisons of the *MOEA* implementations with *Random* search, and of *OMOPSO* with any other *MOEA* implementation, confirmed a statistically significant difference in all performance metrics. The pair-wise comparison of *NSGAII*, *GDE3*, AbYSS and *MOEAD* in the *Additive Epsilon Indicator*, *Contribution*, *Generational Distance*, *Hypervolume*, and *Inverted Generational Distance* performance indicator measures do not feature any noteworthy statistically significant outcomes. The presentation of the p-values has limited value in terms of determining the actual performance difference of the individual *MOEA* implementations. Hence, the reporting of the p-values is omitted here.

## 5. CONCLUSIONS

This paper has described a framework that enables the reconstruction of software architecture configurations on different abstraction levels and considers aspects of the desired conceptual architecture model. This reconstruction is achieved using metaheuristic search algorithms that utilise a number of different software metrics to drive the architecture reconstruction. This dynamic problem representation in combination with the implemented flexible objective configuration approach contribute a framework that can help the user to gain valuable insight into the problem domain of architecture reconstruction and software modularisation. Additionally, the extensible integration of a range of established optimisation libraries now enables the application and performance evaluation of a diversity of MOEA implementations and variation operator tunings in architecture reconstruction application contexts.

The results presented in this paper suggest that many MOEA implementations have the potential to reconstruct the architecture of systems of different complexity. It has been demonstrated in the evaluation that the search converges the architecture configurations towards desired software architecture design metrics. Solutions that feature objective measures, that would be acceptable in practice, in all objectives of the applied objective configuration, could be identified within the smaller software systems that have been considered in this study. Additionally, a range of MOEA performance metrics are presented with the outcome that the sophisticated algorithms perform significantly better than random search. However, the large number of objectives used in this study potentially limit the performance of the algorithms considered. Since the conduct of this research, there has been a growing interest in the use of the NSGA-III algorithm [42] for highly dimensional problems. This algorithm has successfully been applied to a number of software architecture problems [37] and potentially offers significant benefit in supporting the trade-off analysis in the Rearchitecturer tool.

This work is currently limited as a result of its focus on purely quantitate analysis of the algorithm performance and the non-inclusion of a qualitative evaluation of the resulting software systems. Future work will address such a qualitative evaluation involving experienced software architects working with known systems.